# International Promotion Patterns in the Smart City Literature: Exploring the Role of Geography in Affecting Local Drivers and Smart Cities' Outcomes

**Filippo Marchesani, Francesca Masciarelli, and Andrea Bikfalvi**

**Abstract** The rise of smart cities is a significant trend in urban development. However, only in recent years has the focus on the international promotion of these cities become prominent. Despite ongoing academic discussions on the impact of smart city development on urban environments, the global recognition of smart cities is uncertain due to their multidisciplinary nature. Therefore, we conducted a systematic literature review of articles published in top-tier peer-reviewed journals from 2008 to December 2021, providing a comprehensive analysis of existing literature. Specifically, by focusing on the influence of geographical location on cities' international promotion strategies, we highlight the local drivers and the outcomes of smart cities' urban trajectories in different geographical contexts. The paper concludes with a conceptual model aiming to contribute to smart city debates by providing further evidence of the role of geographical location in smart city trajectories.

F. Marchesani (✉)
University of "G. d'Annunzio" Chieti – Pescara, Pescara, Italy

Department of Business Administration and Product Design, University of Girona, Girona, Spain
e-mail: filippo.marchesani@unich.it; f.masciarelli@unich.it

F. Masciarelli
University of "G. d'Annunzio" Chieti – Pescara, Pescara, Italy

A. Bikfalvi
Department of Business Administration and Product Design, University of Girona, Girona, Spain
e-mail: andrea.bikfalvi@udg.edu

## 1 Introduction

The role of smart cities in the urban panorama has gained great scholarly attention in recent decades as it directly influences a plethora of long-run objectives for the current cities, such as integrating and advancing economic, social and environmental areas [1]. Today, smart cities are capable of integrating both innovative models and technologies to optimize local processes and services, influencing the urban environment and playing a catalytic role in the urban landscape to becoming a context capable of attracting and retaining human and financial capital [2–4].

Although both scholars and practitioners believe that smart cities have the potential to shape the urban future, there is still an open discussion on the definition and conception of the smart city and their impact on the national and global level [5, 6]. Today, there is still an unclear conceptualization of the smart cities' implementation, combined with an academic gap on the role of geography in smart cities' initiatives worldwide [7, 8].

This state of art opens up different topics for discussion and emerging perspectives about the correct definition and implementation of smart cities. On the one hand, the implementation of "smart" in cities' development has led cities to a social, cultural, technological and economic environment suitable for citizens and stakeholders [9, 10]. However, on the other hand, it has created a new global competition to attract citizens, tourism, companies and financial capital by offering an internal context that considers its users at the center of urban development [8, 11]. In these trajectories, the needs of cities are different as they must consider the intrinsic development of the local context combined with the conceptualization of "smart" in cities.

Cities are becoming an entrepreneur of themselves and seek to acquire financial, human and institutional resources to reorganize their influence locally [12–15] and elevate their status on the international scene [11]. Therefore, the development and practices of smart cities' openness are essential to bypass the static vision of the city and move into a globalized and competitive context. However, this opening up of smart cities is still in a preliminary but recently has attracted a large stream of attention in the literature [8].

The promotion of cities through international promotion strategies follows the same strategic development process as the one implemented by companies and seeks to involve potential users by offering an urban context suitable for stakeholders, creating real global competition for urban contexts that aim to attract companies, innovative entrepreneurs, new citizens, but also increase tourism, and promote the financial flows [3, 16–18].

This work aims to systematically map the smart cities' literature in that interception by providing a holistic understanding of the development and implementation of smart cities in the global panorama. Specifically, we highlight how worldwide geographical areas adopt different international promotion strategies and development mechanisms driven by structural and strategic motivation (input) and smart cities' outcomes (output) according to their objectives and needs.

Thus, our research questions are: *"What are the relationships between the geographical area and the international promotion strategies of smart cities?"* and *"What are the local drivers and the smart city outcomes in the international promotion of smart cities?"*

To answer this question, our study is structured as follows. First, we delineate the conceptual boundaries useful to outline the research field and define the methodological and structural development of the review. Then, we define our research strategy by outlining the methods and the quality criteria to ensure a systematic and robust approach. The next section provides a descriptive overview of current research, followed by a thematic analysis of the findings. Specifically, in the first part, we highlight the publication trends, information, journal outlets, methodology applied, and research areas. In the second part, we analyze the geographical evolution of publications referring to academic production and the sampling of quantitative and qualitative articles on the subject, highlighting the different development trajectories and discussing the results on multiple levels such as macro-regions, countries, and cities level. Finally, we discuss practical and theoretical implications for academics and policymakers and suggest future research directions.

## 2 Conceptual Boundaries and Methodological Approach

To systematically analyze the literature on the interception between international promotion strategies and the development of smart cities, we adopt the systematic literature review methodology [8, 19]. This approach provides a method to identify, select, analyze, and synthesize existing literature based on a clear and repeatable protocol. We identify robust results and conclusions through an in-depth analysis of the area examined.

To assess this systematic literature review process, we establish the primary conceptual criteria to better outline our research area. In defining the keywords related to the international promotion strategies, we focus on the vision proposed by the international marketing research defined by the Journal of International Marketing that considers the concept of marketing in relation to the international arena where marketing activities occur beyond national and cultural boundaries [8].

Thus, we rely on the definition of marketing proposed by the American Marketing Association (2013): "the activity, set of institutions, and processes for creating, communicating, delivering, and exchanging offerings that have value for customers, clients, partners, and society at large" (AMA 2013). This definition allows us to evaluate the evolution of the studies related to the relationship between cities and various stakeholders from a global competition and development perspective. Cities are opening up and competing globally for human and financial capital, tourism, structural funds, and innovation [11, 20–23]. In this competitive environment, international promotion strategies allow cities to adapt to users and stakeholders in a customized manner, responding to their needs and becoming part of an increasingly competitive context [24]. In the past, local government regulation was considered one of the main obstacles to cross-border development, suggesting that legal restrictions in certain countries dictate modifications in products, prices, and promotional strategies for firms and local development. To survive, cities need to open up beyond the local context and be considered as companies operating in a globally competitive context. Consequently, cities must also outline their internal and international promotion strategies to develop and manage medium/long-term development plans [11].

In defining the keywords related to smart cities, we rely on the definitions of smart cities described in the literature [5] and on a more in-depth review of the reference literature published in the top international journals to select the most appropriate smart cities keywords [8, 25]. Today, the concept of smart cities can be analyzed according to different perspectives that consider different dimensions such as governance, environment, economy, people, living, and mobility [26]. This multidisciplinary area of intervention leads us to consider the city's evolution from a smart perspective based on the different declinations we considered in our set of keywords. In approaching these conceptual boundaries, we only consider international marketing and smart cities concepts, excluding studies that deal with the concept of smart cities but do not address this relationship. For this reason, papers concerning smart cities in terms of internal development are excluded from our sample.

### 2.1 Search Strategy

To deeply explore this research area, we have relied on EBSCOhost's Business Source Premier database as the primary source of research as it provides a solid level of coverage of the major reference journals. This database allows to evaluate the entire selection process thoroughly and reproduce the process by tracing the analysis steps methodically. As a common practice in literature reviews, we performed a three-level search considering keywords on titles, abstracts, and subject terms. Our selection of keywords is based on multiple qualitatively and contextually relevant factors, elaborating on an analysis of the previous literature reviews published in the main international reference journals. We considered the period ranging from January 2008 up to and including December 2021. We used January 2008 as the cutoff starting point, given that the first article matching the keywords appeared in 2008.

The usage of standard Boolean operators allowed for the creation of a single search algorithm. The keyword search algorithm performed by combining the concept of smart city and international marketing strategies was: *("marketing" OR "customer" OR "consumer" OR "export" OR "market" OR "entry" OR "internationalization" OR "internationalization" OR "network" OR "innovation" OR "product" OR "price" OR "promotion" OR "distribution" OR "channel" OR "brand" OR "Internet") AND ("Smart City" OR "Smart Cities" OR "Wired City" OR "Digital City" OR "Information City" OR "Intelligent City" OR*

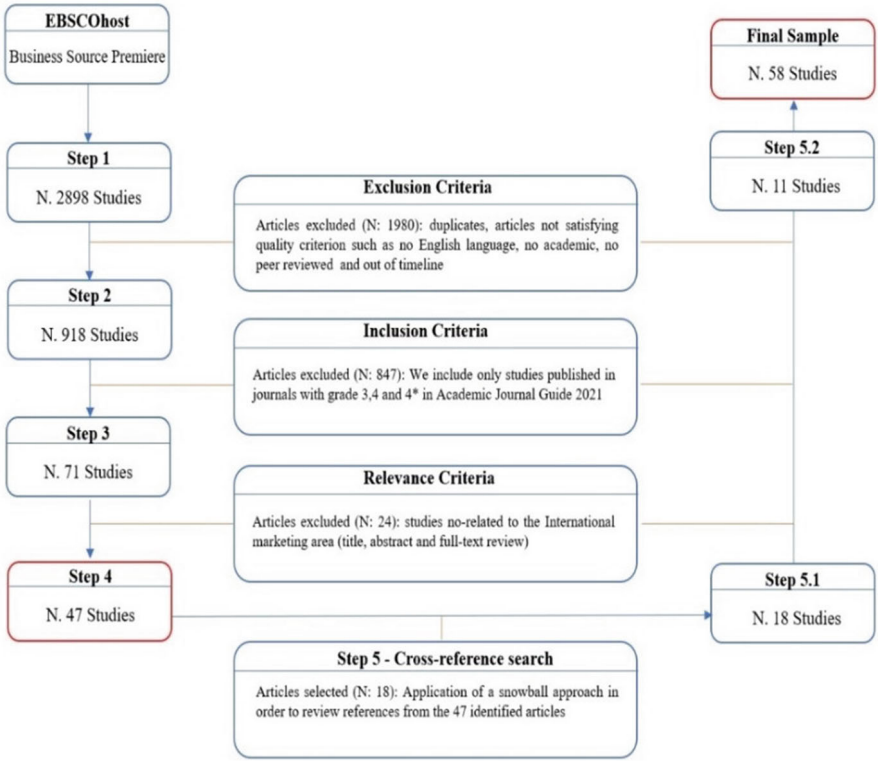

**Fig. 1** Research strategy and boundaries

*"Information City" OR "Information cities" OR "Knowledge-based City" OR "Knowledge City" OR "Knowledge cities" OR "Liveable City" OR "Liveable cities" OR "Ubiquitous City").* In this selection, we included the keywords related to the reference area according to the associated literature in the top journals regarding smart cities [8] and international marketing strategies [27].

After this screening process, 2898 studies relevant to this research area were selected. Subsequently, in order to have a more precise understanding of the role of international marketing strategies in Smart Cities, we decided not to consider studies focused on a national level and focusing only on the international dimension. Our research was thus structured to include studies that focused on the global vision of the smart city and the international marketing strategies (direct and indirect) used by the city to position itself globally. This process allows us to better understand the different dimensions of Smart Cities' development strategies on the international landscape.

Starting from this preliminary analysis, we applied exclusion and inclusion criteria to ensure the high as possible for the sample taken into consideration (See Fig. 1). First, to guarantee and high standard level, we excluded articles according to the quality academic criteria usually adopted in the systematic literature review process. Thus, we exclude book chapters, editorials, conference papers, extended abstracts, book reviews, non-academic articles, and no peer-reviewed articles [28, 29]. Second, we excluded articles not available in English. We adopt this research strategy to focus only on the scientific knowledge base represented by the majority of scientific journals [30] to assess the review process systematically. This exclusion criterion narrowed our sample to 918 potentially relevant studies.

Subsequently, we applied the quality criterion to further narrow the sample under consideration in order to ensure the quality of studies included in their review. We limited our review to studies in peer-reviewed journals ranked 3, 4 or 4 * in the Academic Journal Guide 2021 (previously considered ABS2018) journals list, which includes only top-tier academic journals, being this a frequently adopted method for identifying scholarly debates and research trends when performing a literature review of given a research area [29, 31]. This additional screening led the total sample to 71 potentially relevant studies.

At this point, once the final sample was identified, we moved on to the reading of every single paper starting from the abstract and possibly the entire article to evaluate the contribution from an international perspective of the selected articles. This further control led to the exclusion of 24 articles bringing the final sample to 47 articles. Finally, to ensure that no articles were missed, we compute the cross-reference analysis also adopted in other SLR [32] by applying the backward and forward snowballing method to search into the deep of our sample. Specifically, we singularly examined the reference lists of all articles selected in the review process. This additional step identifies 18 other articles (Step 5.1) that have undergone the previous evaluation process through inclusion, exclusion and quality criteria. In conclusion, 11 articles met the selection criteria (Step 5.2) added to the previously selected sample, bringing the final sample to 58 articles, as reported in Fig. 1.

Considering this final sample, we constructed an extraction table that considered the main details of the articles, such as the institution of author affiliation, methodology, sample, geographical information, and any relevant information such as theoretical and main results, practical implications, and future research to capture the geographical development of this research area. This allows us to systematically analyze the final sample and extract interpretive elements from it.

## 3 Results

### 3.1 Descriptive Analysis

Since 2008, the number of articles published every year had an increasing trend, as it can be observed in Fig. 2, with a generalized peak in the last 2 years that has shown an increasing trend compared to the recent development (2020 = 7 and 2021 = 12), highlighting the fact that is still growing and to a rapid degree. Specifically, the results show that academic output on the topic has increased exponentially over the past decade following the smart cities trajectories that have begun to move from a purely national context to a more global vision [33]. Moreover, it emerges that 42% of the scientific production of the last 13 years has been disseminated in publication outlets in the past 2 years, showing that the reviewed research area is quickly shifting into a more advanced stage. As we will also highlight later, this research has not remained anchored to the internal development of the city concept but is evolving, including practices and strategies that are ever closer to a competitive and managerial context.

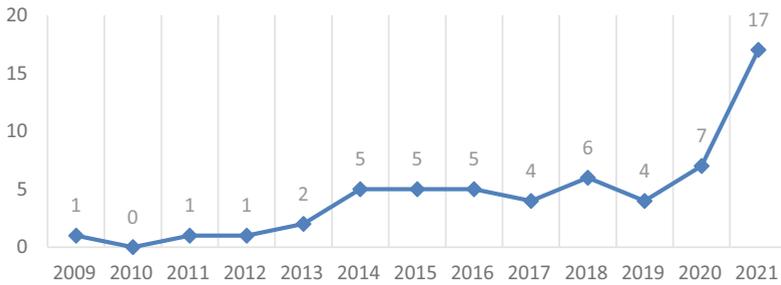

**Fig. 2** Evolution graph of the distribution of the number of publications per year

The journals considered in this area are Urban Studies with 16 articles and Government Information Quarterly with 12 articles, covering more than 50% of the total sample. It is important to focus on this preponderance of articles because the sample shows different trends. On the one hand, there is a constant evolution from 2016 to 2020 of publications in Government Information Quarterly, mainly focused on governance and development policies considering theoretical and qualitative articles. On the other hand, a trend consolidated over the years in Urban Studies with a peak in the last year mainly focused on urban adaptation and development in cities. Other journals of note include Cambridge Journal of Regions, Economy and Society with 6 articles, California Management Review (3) and European Urban and Regional Studies, Technovation, Production Planning and Control and Public Management Review appear with 3 articles. The rest of the scientific production is distributed in different journals and thematic areas, as it can be interpreted from the text in the next paragraphs (See Table 1).

The thematic areas that emerge from our sample are various, as well as the conjunctions of this research area from a methodological perspective. However, consistently with the smart city concept, the emerging reference areas mainly concern urban study and the study of systems that support data-intensive applications. Table 2 show that the predominant subject area in this study is Regional Studies, Planning and Environment, which covers almost 50% of the total sample with 23 articles, followed by Information System with 12 articles. Other publications concern, Innovation (5), General Management, Ethics, Gender and Social Responsibility (3), Public Sector and Health Care (3), Operations and Technology Management (3), Operations Research and Management Science (2), Marketing (1) and Economics, Econometrics and Statistics (1). Deeply analyzing the final sample of the literature, several significant methodology-related trends emerge that evidence that the field is at an early stage and that the current advancement of this research trend is more theoretical than practical. This is confirmed by the study of applied methodologies which presented a large number of theoretical contributions (20) specifically referred to the first part of the timeline table and highlights how the study of the phenomenon has been addressed in this perspective first at a theoretical level and subsequently implemented in case studies and practical.

**Table 1** Articles distribution in Academic Journals

| Articles | Journals | AJG ranking |
|---|---|---|
| 1 | Research Policy | 4* |
| 3 | Public Management Review | 4 |
| 1 | Production and Operations Management | 4 |
| 1 | Journal of Supply Chain Management | 4 |
| 16 | Urban Studies | 3 |
| 12 | Government Information Quarterly | 3 |
| 6 | Cambridge Journal of Regions, Economy and Society | 3 |
| 5 | California Management Review | 3 |
| 3 | European Urban and Regional Studies | 3 |
| 3 | Technovation | 3 |
| 2 | Production Planning and Control | 3 |
| 1 | IEEE Transactions on Systems, Man, and Cybernetics | 3 |
| 1 | Energy Economics | 3 |
| 1 | International Marketing Review | 3 |
| 1 | IEEE Transactions on Systems, Man, and Cyb. | 3 |
| 1 | Technological Forecasting and Social Change | 3 |

**Table 2** Methodology and subject area

| | Methodology | | | | | |
|---|---|---|---|---|---|---|
| | Theor. | Empirical | | | Review | Total |
| Subject areas | | Quantitative | Qualitative | Mixed-methods | | |
| Econ, econometrics and statistics | | 1 | | | | 1 |
| General management, ethics, gender and social responsibility | | | 3 | | | 3 |
| Information systems | 5 | 2 | 5 | 1 | 1 | 14 |
| Innovation | 2 | | 2 | | 1 | 5 |
| Marketing | | | | | 1 | 1 |
| Operations and technology management | 3 | | | | | 3 |
| Operations research and management science | 1 | | 1 | | | 2 |
| Public sector and health care | | | | 2 | 1 | 3 |
| Regional studies, planning and environment | 9 | | 16 | 1 | | 26 |
| Total | 20 | 3 | 27 | 4 | 4 | 58 |

This is confirmed by the study of applied methodologies which presented a large number of theoretical contributions (20) specifically referred to the first part of the timeline table and highlights how the study of the phenomenon has been addressed in this perspective first at a theoretical level and subsequently implemented in case studies and practical. Moreover, great attention is concerned with the number of studies using a qualitative methodology approach. In particular, considering 34 empirical studies in our sample, 27 use a qualitative approach. Based on this qualitative contribution, a large part presents single or multi-case studies to analyze the phenomenon within a specific urban context or a nation. These results reveal how this phenomenon is still in an embryonic phase where various concepts, objectives and developments are still at an embryonic stage and not well fixed in the literature. The remaining empirical studies apply a mixed-methods approach (4 articles), and only 3 articles apply a quantitative approach.

Thus, the evolution of this research topic lacks an in-depth analysis and exploration in practices, an issue that future academics and practitioners should converge on and solve, as a profound analysis of the phenomenon will allow to build a strong theoretical foundation and further enhance the research area in theory and practices. The presence of 4 "review" articles related to governance [34–36] and marketing [8, 37] and highlights how there is a need to re-examine the current static and urban vision of the city into a (more) dynamic and global perspective.

## 3.2 Geographical Distribution and Analysis

The geographical analysis, placed at the centre of this systematic literature review, highlights important conjectures on the intersection between international marketing strategies and smart cities. Indeed, the expansion of cities outside the urban context is a phenomenon that mainly interests certain geographic areas. Below, we begin with an analysis of the authors, the geographical context, and the institution of the authors' affiliation. Then, we analyze the sample in-depth, evaluating the countries involved and the cities considered in the selected research.

Starting from the reference sample, we analyze the authorship of the publications in question, focusing on the number of authors, nations, and institutions to highlight the multi-project quality and cross-knowledge of the research. The number of authors is in line with the general academic context of reference, where most publications have 3 or more authors. In contradiction with this background, it is interesting to comment on the nature of these collaborations. The sample highlights that although the number of institutions involved is balanced (see Table 3), the reference nations involved in the various publications are strongly unbalanced to a single country (34 Articles) compared to the plurality of countries involving two (15 Articles) or three countries (9 Articles). This data contrasts with the study of international openness and strategies to promote the city at the international level.

**Table 3** Authorship characteristics per authors, countries and institutions

|               | Numbers of authors (%) | Number of countries (%) | Number of institutions (%) |
|---------------|------------------------|-------------------------|----------------------------|
| One           | 20.75                  | 58.49                   | 35.84                      |
| Two           | 28.3                   | 26.41                   | 30.18                      |
| Three or more | 50.94                  | 15.09                   | 33.96                      |

**Table 4** First authors distribution

| No. | Country        | Macro-regions                |
|-----|----------------|------------------------------|
| 11  | United States  | North America                |
| 9   | United Kingdom | Europe and Central Asia      |
| 8   | Spain          | Europe and Central Asia      |
| 6   | Italy          | Europe and Central Asia      |
| 4   | Canada         | North America                |
| 2   | China          | East Asia and Pacific        |
| 2   | Ireland        | Europe and Central Asia      |
| 2   | Netherlands    | Europe and Central Asia      |
| 2   | Australia      | Europe and Central Asia      |
| 2   | Chile          | Latin America and Caribbean  |
| 2   | UAE            | Middle East and North Africa |
| 1   | India          | South Asia                   |
| 1   | Korea          | East Asia and Pacific        |
| 1   | Kuwait         | Middle East and North Africa |
| 1   | Denmark        | Europe and Central Asia      |
| 1   | Switzerland    | Europe and Central Asia      |
| 1   | Cyprus         | Europe and Central Asia      |
| 1   | Sweden         | Europe and Central Asia      |
| 1   | South Africa   | Sub-Saharan Africa           |

However, the relatively young field of research and methodological developments (as seen above) suggests that in the future, the number of empirical research involving two or more countries will grow exponentially, as will the increase in the phenomenon studied.

Subsequently, by investigating the author's affiliations, we analyze this phenomenon's nature and interest worldwide following analyses at the national and macro-regions level. As it can be observed in Tables 4 and 5, scientific production on the subject is concentrated in Europe and Central Asia with 33 articles (57%) and North America with 15 articles (26%). In detail, we note that European scientific production is mainly developed in 3 countries (UK 9, Spain 8, and Italy 6), which alone cover the majority of the European scientific production on the subject. Moreover, Concerning North America, the publications are divided into 11 USA and 4 Canada. The rest of the scientific production originated in the remaining regions, specifically 3 in East Asia and Pacific (of which two in China and one in Korea), 3 in the Middle East and North Africa (2 in UAE and 1 in Kuwait), 2 in Latin America and Caribbean, 1 in Sub-Saharan Africa and 1 in South Asia. These results show how the academic interest in developing the smart city in a global landscape refers mainly to countries based on a vision of internationalization and openness to the outside world. While the strong technological component would suggest a greater development of these practices in Asia, with cities historically characterized by a robust implementation of big data and smart solutions [38], academic research on the development and promotion of the city in this intersection mainly concerns a trend of research in liberal and capitalist countries. A very particular case concerns the UK, which on the one hand, represents the main producer of academic articles on the subject (see Table 4); on the other hand, it is not strongly represented in the study sample (see Table 6), especially in the recent trend. This trend may be influenced by the "Brexit" that has influenced the trajectories of the development of smart cities from a global perspective, widely represented in our sample in the pre-Brexit period [39], into a more urban and internal perspective.

**Table 5** First authors' distribution per macro-regions

| No. | Region |
|---|---|
| 35 | Europe and Central Asia |
| 13 | North America |
| 3 | East Asia and Pacific |
| 3 | Middle East and North Africa |
| 2 | Latin America and Caribbean |
| 1 | Sub-Saharan Africa |
| 1 | South Asia |

**Table 6** Sample geographical location of empirical papers per country, region and income group

| No. | Country | | Region | Income group |
|---|---|---|---|---|
| 5 | Italy | ITA | Europe and Central Asia | High income |
| 4 | Spain | ESP | Europe and Central Asia | High income |
| 4 | United States | USA | North America | High income |
| 4 | UK | GBR | Europe and Central Asia | High income |
| 3 | Canada | CAN | North America | High income |
| 1 | Austria | AUT | Europe and Central Asia | High income |
| 2 | Brazil | BRA | Latin Am. and Caribbean | Up/Md income |
| 2 | Chile | CHL | Latin Am. and Caribbean | High income |
| 1 | China | CHN | East Asia and Pacific | Up/Md income |
| 1 | Colombia | COL | Latin Am. and Caribbean | Up/Md income |
| 1 | Finland | FIN | Europe and Central Asia | High income |
| 1 | Germany | DEU | Europe and Central Asia | High income |
| 1 | Netherlands | NLD | Europe and Central Asia | High income |
| 1 | India | IND | South Asia | Lo/Md income |
| 1 | Portugal | PRT | Europe and Central Asia | High income |
| 1 | South Korea | KOR | East Asia and Pacific | High income |
| 2 | Sweden | SWE | Europe and Central Asia | High income |
| 1 | Taiwan | TWN | East Asia and Pacific | High income |
| 1 | Philippine | PHL | East Asia and Pacific | Lo/Md income |
| 1 | UAE | ARE | Midd. East and Nrt. Africa | High income |

**Table 7** Sample macro-regions distribution

| No. | Region |
|---|---|
| 20 | Europe and Central Asia |
| 7 | North America |
| 5 | Latin America and Caribbean |
| 4 | East Asia and Pacific |
| 1 | South Asia |
| 1 | Middle East and North Africa |
| No. | *Income group* |
| 32 | High income |
| 4 | Upper middle income |
| 2 | Lower middle income |

Finally, we analyze the geographical distribution of the final sample and the cities considered as case studies. As Table 7 illustrates, the distribution is strongly unbalanced towards Europe (20) and North America (7), followed by Latin America (5) and finally by East Asia and Pacific (4), South Asia and the Middle East (1) and North Africa (1). The nations mainly considered in this field are Italy (5 articles), followed by Spain, the United States and the UK with 4 articles. This evidence may lead to different views. The first is the need to improve the collaboration between practitioners and academics to corroborate the geographical need, national policies and smart cities implementation. The second refers to the current policies related to the countries' openness regarding open data on smart cities' implementation and management. This transparency in management and investment allows for a detailed analysis of development strategies and trajectories, leading to a bilateral relationship between internal development and cities' international promotion.

It is also interesting to note that Latin America and Caribbean countries (5) also emerge in the sample, undergoing strong development from a smart point of view in recent years. These countries are surfing the smart city wave to solve internal problems and join a context that is not yet fully developed, as the case of Santiago de Chile presented in our final sample (see Table 8), which is to take a leap forward and catch up with more developed countries in terms of energy efficiency and environmental care [40, 41]. Moreover, to understand the level of development of the countries analyzed, we code our geographical sample using a three-group country classification that distinguishes between developing, emerging, and developed countries based on the World Bank's grouping of countries relative to annual gross national income (GNI) per capita. Accordingly, the developing group reunites low-income countries; the emerging group comprehends middle-income countries, and the developed category includes high-income countries. We decided to use the World Bank's classification based on the GNI per capita as it is widely used analytically to evaluate the country's development. This country coding allows us to rigorously determine the development of the countries examined by highlighting two considerations. The first is that the vast majority of the countries analyzed concern countries with a high income (32 articles), a sign that promoting the smart city in the global panorama is an elite concept involving more developed countries. The second is that the nature of publications in Upper middle income (4 articles) and Lower middle income (2 articles) countries is relatively recent, highlighting how this phenomenon motivates and pushes even less economically developed nations to draw on this phenomenon from a perspective of international development and promotion. This means that following or reproducing smart city models oriented to promoting the smart city into the global context in developed countries also serves as a model for less developed countries that aim at rising in the urban and geographical context of reference by boosting themselves in the urban environment and raising their status in the geographical area [8].

Table 8 Cities considered in the case study

| Sample City | |
|---|---|
| Code | City |
| ESP | Barcelona (3) |
| ITA | Milan (2) |
| BRA | Rio De Janeiro (2) |
| CHI | Santiago de Chile (2) |
| USA | Seattle |
| CAN | Vancouver |
| AUT | Vienna |
| GBR | London |
| DEU | Cologne |
| NLD | Amsterdam |
| PRT | PlanIT Valley |
| KOR | Songdo |
| SWE | Stockholm |
| TWN | Taipei |
| ITA | Naples |
| ARE | Dubai |
| PHL | Manila |
| CHI | Medellin |
| USA | Annapolis |
| PHI | New Clark City |
| CAN | Iqaluit |
| CAN | Nunavut |

The city-level represents a large percentage of the empirical studies analyzed. Specifically, analyzing the sample of selected articles, 27 cities were used as case studies in the research. According to the sample, the main cities analyzed concern two European cities, Barcelona (3) and Milan (2), and two Latin American cities, Rio de Janeiro (2) and Santiago de Chile (2).

The precursor cities of this concept are mainly located in Europe. Barcelona (3 articles) has become a benchmark Smart City in the twenty-first century. Many studies start right from Barcelona to extend theories of reference globally, as to date, international competitiveness is driven by the innovativeness of cities [42]. Barcelona is a forerunner in this sector as it can provide a strongly smart context for the development and attraction of businesses and technology-based clusters [1, 43].

While European cities were precursors of the "smart" development of the city, it is interesting how cities such as Rio de Janeiro and Santiago de Chile are among the main cities analyzed. In this sense, cities in Latin America are striving to become global. According to Roy [44], the evolution toward a "smart" city in Latin America lies in the melting pot of development modernization projects, immigration, and neoliberal governance, which has led to a more global vision in recent years [40].

On the contrary, few projects concern the Asian continent, countries that have always been ahead of the smart city development but are not marginally included in the international promotion of the city. This evidence is configured in the different development trajectories of smart cities and will be discussed later to highlight a critical and theoretical framework of state of the art.

## 4 Discussion

This systematic literature review aims to characterize a link between smart cities and international promotion strategies in the global landscape. This procedure allows us to capture the direction of the link and the emerging geographical pattern in this relationship. The following conclusions result from this review. Large bodies of academic literature address smart cities and cities' development from an international perspective. However, literature documenting a clear linkage between the two is scant at the same time. In the intersection between international promotion strategies and smart city development, the geographic location plays a fundamental role by influencing internal drivers in developing smart cities, urban growth and development strategies at a national and international level. For example, if in certain geographic areas there is a "smart" implementation based on the citizen-centric perspective [9, 45], in other geographic areas, even they are considered "at the global vanguard" in smart city development [46, p. 71], the approach is mainly top-down (i.e. South Asia). Consequently, strategies and cities' development will differ according to their geographical and structural characteristics.

The lack of a direct connection in the literature is given not only by the (relative) recent development of the smart city concept [47] but also by a different vision of the same concept according to the geographical environment. This evolution leads governments and policymakers to approach the implementation of the smart city differently depending on the reference context and development objectives. In this article, we highlight how the geographical context influences the development trajectories of cities in relation to international promotion strategies (see Fig. 3). Specifically, it highlights how different inputs in the development of cities lead to substantial and well-defined outputs, which turn into different smart city trajectories linked to the geographical reference area. Thus, this review highlights three different approaches to the international promotion of cities: (1) the global-promotion approach of cities located in Europe and North America, which aim to compete worldwide for innovation, tourism, and human and financial capital; (2) the emerging approach of cities in Latin America which aim to raise their status in the reference context and (3) the top-down approach of cities in East and South Asia focusing on urban advancement rather than global promotion.

| Macro-regions | Europe and North America | Latin America | East and South Asia |
|---|---|---|---|
| Local Drivers | • Technology-based and "smart" transition<br>*(Berrone, 2016; de Falco, Angelidou, & Addie, 2019; Leitheiser & Follmann, 2020; Vanolo, 2014)*<br>• Sustainability and efficiency<br>*(Herrschel, 2013; Neumann, 2019; Paskaleva & Cooper, 2018; Valdés, Cook, & Potter, 2018)*<br>• Citizen-centric implementation<br>*(Gagliardi et al., 2017; Leitheiser & Follmann, 2020; Nesti & Graziano, 2020; Neumann et al., 2019; Paskaleva, 2018)* | • Urban condition and "smart" implementation<br>*(Almirall et al., 2016; Hollands, 2015; Jirón, Imilán, Lange, & Mansilla, 2021; Muñoz & Cohen, 2016)*<br>• Urban efficiency advancement<br>*(Goodspeed, 2015; Irazábal & Jirón, 2021; Muñoz & Cohen, 2016; Sancino & Hudson, 2020)*<br>• Citizens-based implementation<br>*(Goodspeed, 2015; Irazábal & Jirón, 2021; Jirón et al., 2021)* | • Modelling urban architecture<br>*(Carvalho, 2015; He et al., 2020; Mouton, 2021; Zhang, Zhao, & He, 2020)*<br>• Technology-based development<br>*(Carvalho, 2015; Chatterjee et al., 2018; He et al., 2020; Mouton, 2021)*<br>• Government-based implementation<br>*(Chang, Jou, & Chung, 2021; Chatterjee, 2018; He, Yang, Bai, & Zhang, 2020; Mouton, 2021)* |
| Smart City Outcome | • "Smart" advancement and citizens engagement<br>*(Alcaide-Muñoz, 2017; Gagliardi et al., 2017; Kitchin, 2015; Paskaleva & Cooper, 2018)*<br>• Cities' attractiveness<br>*(Betz, Partridge, & Fallah, 2016; Christofi, 2021; Hollands, 2020; Buck & While, 2017)*<br>• Competitive urbanism<br>*(Almirall et al., 2016; Herrschel, 2013; Kitchin, 2015; Leitheiser & Follmann, 2020)* | • Enhance the structural eco-system<br>*(Irazábal & Jirón, 2021; Muñoz & Cohen, 2016; Sancino & Hudson, 2020)*<br>• Tackle urban problems<br>*(Goodspeed, 2015; Hollands, 2015; Irazábal & Jirón, 2021; Jirón et al., 2021)*<br>• Enhance their status in the geo-local area<br>*(Goodspeed, 2015; Jirón et al., 2021; Muñoz & Cohen, 2016; Sancino & Hudson, 2020)* | • Urban advancement and user-connection<br>*(Chang et al., 2021; Chatterjee et al., 2018; He et al., 2020; Mouton, 2021)*<br>• Socio-technical advancement<br>*(Carvalho, 2015; Chang et al., 2021; Chatterjee et al., 2018; Zhang et al., 2020)*<br>• Information system and monitoring<br>*(Carvalho, 2015; He et al., 2020; Zhang et al., 2020)* |
| Approach | *Global-promotion approach* | *Emerging approach* | *Top-down approach* |

Fig. 3 Conceptual model

The first approach concerns the cities located in Europe and North America, which compete globally and base their strategies on a capitalist perspective to attract tourism, human and financial capital, and companies. The smart evolution in those areas is driven by the technological and smart transition in services and processes that aim to improve the development of cities by placing the citizen and users at the centre of cities' strategies [48–50]. At the same time, the current need to be sustainable and efficient in medium-long term development leads cities to start and invest in this transformation process [16]. This evolution has impacted both internal and external perspectives and aims to evolve the static concept of cities towards a more comprehensive and global vision [6]. On the one hand, a global vision in cities development makes the city attractive to citizens and stakeholders and, on the other, enhances its status worldwide, useful for attracting human capital, talents and companies to compete locally and internationally [8, 51, 52].

The second approach focuses on the emerging trend in Latin American countries. These countries recently promoted smart city projects by initiating a "twinning relationship" with North American and European smart city projects (i.e., Rio, Santiago and Medellín whit Barcelona) to re-shape their own cities' trajectories due to the necessity to tackle internal problems in terms of a social and urban environment [40, 53]. The thrust of this transformation is given by the need to improve urban efficiency and put the citizen at the centre of this urban renewal project [54, 55]. In this case, the desire/need to reorganize the local context is the basis for developing their strategies nationally and internationally to enhance their position in the regional eco-system, tackler the urban issues and promote their status in the reference context [41, 54].

Finally, the most veiled result reveals the relatively low presence of studies concerning east and south Asia. Specifically, many papers dealing with the topic of smart cities in Asian regions do not appear in our sample as they have a top-down approach [56]. The reason is that the smart cities of these regions focus on technology and control tools such as big data to manage their development trajectories [38]. The final aim becomes internal management rather than international promotion by privileging socio-technical advancement and enhancing the connectivity and monitoring of its users [57, 58].

Thus, we assume that European and North American smart city initiatives have a more market-oriented and global vision. This vision impacts the development decisions of urban contexts and the cities themselves, even in opposition to city governance in east and south Asia, in which development extends from governance to stakeholders. In this context, the initiatives in the cities of Latin America lie between the two previous approaches. On the one hand, they aim to solve internal problems through the evolution of the city from a smart perspective. On the other, they plan to promote their position on the local and international scene.

## 5  Conclusion

In conclusion, the study, in addition to summarizing the geographical and urban contents on the specific topic through the approach of the SLR, systematically documents the recent research trends and the trajectories of the cities at the geographic level, showing different approaches to the concept of the smart city due to different technological, environmental, social, and evolutionary components.

The current results can be useful for both academics and practitioners from a strategic perspective. Specifically, from a theoretical view, we integrate new knowledge on the topic and support academics about the role of smart city initiatives in the urban environment. Furthermore, the different cities trajectories in morphologically, culturally, politically, and economically different countries influence the approaches and objectives of smart city projects, which can open up new theoretical developments on the practical evolution of smart cities from an international perspective [57]. Thus, we also map the literature on the global promotion of the city, providing new insights and theoretical evidence at a geographical level. An in-depth understanding of the geographical distribution would help the academics design a solid framework to promote and implement smart cities and generate a new flow of knowledge in the interplay between cities promotion, geographical area, and urban trajectories.

Finally, from a practical perspective, we support governance and policymakers in highlighting how different trajectories could modify the existing scenarios, reshaping a context that today is strictly focused on the operational objective of the smart city [59]. This perspective can offer insights into international promotion strategies and strengthen the current development trends in the different geographical areas highlighted in the literature.

Furthermore, these results open new research on the role of city promotion in the global context. The reduced number of empirical studies and the recent development of this research trend open to various studies on the subject. These studies can be traced horizontally by analyzing the reference context and the inputs and outputs required depending on the geographical area or vertically investigating the causes at the political and social levels. The international opening concert of smart cities continues to be underdeveloped. This feeds new lines of research on internal drivers and expected benefits (smart city outcome) in the international landscape.

To conclude, there are limitations that need to be acknowledged, consistent with all systematic literature reviews, which are related to the selection of the sample and the ongoing discourse within the references. Firstly, our choice of keywords and the utilization of EBSCOhost may have led to the inadvertent omission of pertinent literature. Secondly, our scope was centered exclusively on the international facet of the research, leading to a comprehensive analysis of the selected sample.